\documentclass[acmtog,timestamp]{acmart}
\acmJournal{TOG}
\setcopyright{none}
\usepackage{amsmath, amsthm, amsfonts, amssymb}
\usepackage{graphicx}
\usepackage{wrapfig}
\usepackage{subfig}
\usepackage{color}
\usepackage{xspace}
\usepackage{overpic}
\usepackage{subfig}
\usepackage{wrapfig}
\usepackage{enumitem}
\usepackage{mathrsfs}
\usepackage{bm}
\renewcommand{\textrightarrow}{$\rightarrow$}

\graphicspath{{figures/}}

\definecolor{turquoise}{cmyk}{0.65,0,0.1,0.1}
\definecolor{purple}{rgb}{0.65,0,0.65}
\definecolor{dark_green}{rgb}{0, 0.5, 0}
\definecolor{orange}{rgb}{0.8, 0.6, 0.2}
\definecolor{red}{rgb}{0.8, 0.2, 0.2}
\definecolor{brown}{rgb}{0.5, 0.16, 0.16}

\newcommand{\uri}[1]{{\color{black}#1}}

\setcitestyle{none}

\begin{document}

%\title{Learning Material Constitutive Models from Captured Sparse Data}
%\title{Neural Material: Data Driven Material Constitutive Models} 
%\title{Neural Forces for Material Constitutive Models}
%\title{Neural Material: Learning  Constitutive Models from  Sparse Data}
\title{Neural Material: Learning Elastic Constitutive Material and Damping Models from  Sparse Data}

\author{Bin Wang}
\affiliation{\institution{Beijing Film Academy}}
\email{binwangbuaa@gmail.com}

\author{Paul Kry}
\affiliation{\institution{McGill University}}
\email{kry@cs.mcgill.ca}

\author{Yuanmin Deng}
\affiliation{\institution{Shandong University}}
\email{yuanmin.deng@gmail.com}

\author{Uri Ascher}
\affiliation{\institution{University of British Columbia}}
\email{ascher@cs.ubc.ca}

\author{Hui Huang}
\affiliation{\institution{Shenzhen University}}
\email{hhzhiyan@gmail.com}

\author{Baoquan Chen}
\affiliation{\institution{Beijing Film Academy / Peking University}}
\email{baoquan.chen@gmail.com}

\renewcommand\shortauthors{}	

\begin{abstract}

{\color{black}{The accuracy and fidelity of deformation simulations are highly 
%depend on 
\uri{dependent upon} the underlying constitutive material model. Commonly used linear or nonlinear constitutive material models only cover a tiny part of 
%entire possible material space.
\uri{possible material behavior}. 
	In this work we propose a unified framework
	\uri{for modeling deformable material.}}}
%	 to model material in the space.} }
%
The key idea is to use a neural network to correct a nominal model of the elastic and damping properties of the object. The neural network  
encapsulates a complex function that is hard to explicitly model. 
%and 
\uri{It} injects
{\em force corrections} that help the forward simulation to more accurately predict the true behavior of a given soft object, which includes non-linear elastic forces and damping.
{\color{black}{
%To mostly 
\uri{Attempting to} satisfy the requirement from real material interference and animation design scenarios,} }
we learn material models from examples of dynamic behavior of a deformable object's surface. The challenge is that such data is sparse as it is consistently given only on part of the surface. 
Sparse reduced space-time optimization is employed to gradually generate increasingly
accurate training data, which further refines and enhances the neural network. 
We evaluate our choice of network architecture and show evidence that the modest amount of training data we use is suitable for the problem tackled. Our method is demonstrated with a set of synthetic examples.
%, as well as by using data captured from real world homogeneous elastic objects.
% PGK needed to supress all mention of captured data.

\end{abstract}

%\begin{abstract}
%We learn material models from captured examples of dynamic behavior of an object's surface. We call this sparse data because it is only surface data and only part of the surface.  Our material model consists of a nominal principal stretch based finite element model, which we correct with a neural network to provide non-linear elastic forces and damping.  \pk{DESCRIBE MORE...}.  We evaluate our choice of network architecture.  We show evidence that the amount of training data we use is suitable for the problem.  We demonstrate our  method with a set of synthetic examples, as well as using data captured from real world homogeneous elastic objects.
%\input{abstract}
%\end{abstract}


%\ccsdesc[300]{Computing methodologies~neural networks}


\begin{teaserfigure}  %% linewidth set .75 simply to have intro finsh on page 1
	\includegraphics[width=.99\linewidth]{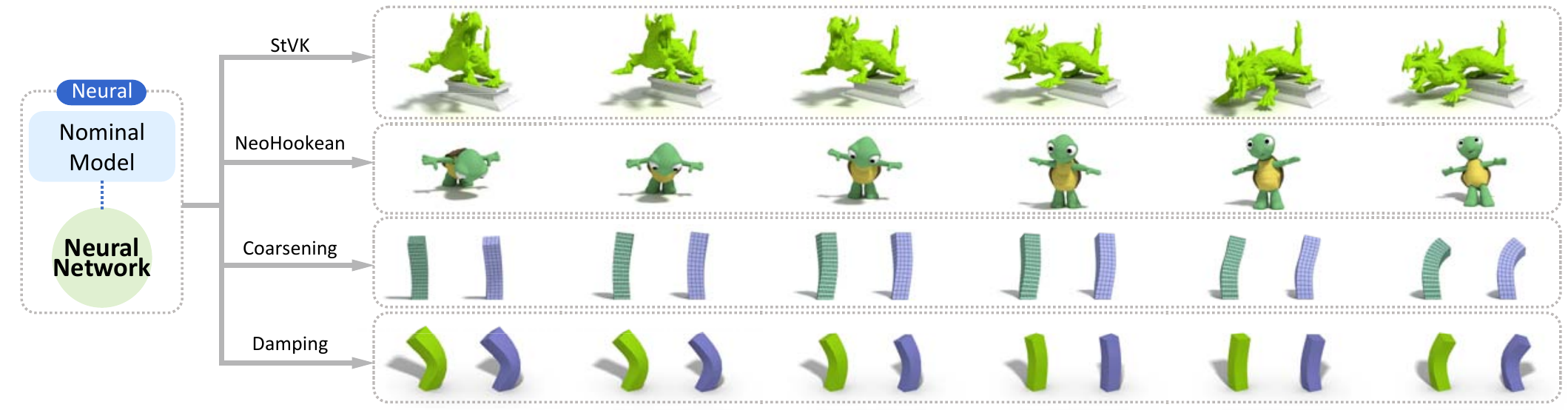}%{tesear.png}
	\caption{Our neural material model learns a correction to a nominal material model that
		allows us to accurately capture nonlinearity of different constitutive material models, realize 
		deformable mesh coarsening, and model damping effects implicitly.}  
	\label{fig:tease}
\end{teaserfigure}

\maketitle

\section{Introduction}
\label{sec:intro}

The simulation and calibration of deformable objects is ubiquitous in computer graphics and robotics research due to the
large number of varied applications in which such tasks naturally arise.
These include animation, movie making, medical treatment, and manufacturing.
Impressive and demanding physics-based animations have almost become routine.

The demand for accurate simulations has in turn highlighted the need to better capture the actual constitutive model
associated with a given soft body under deformation, as it now affects more directly the resulting simulations.
To avoid manual tuning, data driven methods can be deployed. For instance, advanced scanning and sensing technology can be used to faithfully capture a deformation behavior under external force, and used to estimate the parameters of a mathematical model. 
However, common simulation approaches use simple or approximate constitutive models 
that often fail to capture the desired behavior of real objects, fine level simulation of heterogeneous models, or artist examples.

%Physics- based deformation simulation is widely used in many region, such as animation, movie making, medical treatment, and manufacturing. These simulations require mathematical models which concisely describe the physics of the deforming material. Typically, to achieve the desired effect, some manual, tedious and time consuming tuning of parameters is required by experts. Still, such manual tuning can only approximate the real physical behavior.

Typical simulations involve a constitutive model that contains two force contributions: elastic and damping. 
There are several elastic force models that employ a nonlinear stress-strain relationship, 
for instance extending a linear Hookean regime. 
%as in Figure \uri{2}.
But they are all known to have limited ranges of applicability, especially for large deformations~\cite{Sifakis2012,ciarlet88}.
% PGK: Need to fix as the NN will also have limited range of applicability
For damping, many use the Rayleigh model, but it can be inadequate for visual purposes~\cite{xu2017}. Indeed there is no agreed upon
damping model in the mechanical engineering literature.

%%%% PGK: INSERT TEASER HERE??  BUT ALSO NEED TO REFERENCE IT 
%\begin{figure}  %% linewidth set .75 simply to have intro finsh on page 1
%	\includegraphics[width=.99\linewidth]{Representative.pdf}
%	\caption{Our neural material model learns a correction to a nominal material model that
%		allows us to accurately reproduce the captured trajectory, even when
%		the nominal model differs significantly.}  
%		%Additional examples and  convergence properties of our method can be seen in the supplementary video.}
%	\label{fig:tease}
%\end{figure}

%More specifically, the descriptive power of empirical models of soft material are necessarily simplified, for example, using Rayleigh damping model to allow affordable numerical analysis. Thus, empirical models cannot faithfully reconstruct the true property of the target subject, and they remain limited to describe only a rather small subspace of constitutive material of real objects and deformation behaviors. 

%To extend the scope of empirical models, in 
The goal of this work is to augment the performance of empirical force models using a neural network. 
Neural networks are capable of learning latent complex nonlinear relationships when trained with large amounts of data. 
We leverage on their strong function representative ability as means to compensate for the complex constitutive materials (observed in fine heterogeneous simulations, real scanned trajectories, or artist examples) 
%of real objects 
in the context of a forward simulator. 
The key idea is that a neural network can encapsulate a complex function that is hard to explicitly model. 
The neural network injects ``force corrections'' that help the forward simulation to achieve accurate results.

The training of a neural network to assist in augmenting an empirical model is challenging because there may be no available training data, and example trajectories of desired behavior may be sparse, such as captured by incomplete surface scans.
%and captured data is only sparse and partial. 
Moreover, the data is unlabeled in the sense that there may not typically be any knowledge of the desired material properties of the available data. 
%captured data. 
To alleviate these difficulties, we propose a 
%reinforcement learning strategy 
% PGK: Can this really be considered an RL stategy?  Seems just like a buzz word.  Cleaned this up.
to learn complex constitutive material from 
%only captured 
sparse motion trajectories. 
%The idea is to 
We use a sparse reduced space-time optimization to gradually generate increasingly accurate training data, which further refines and enhances the neural network.
% further and further. 
The basic unit that computes the forces in our simulation has contributions from both a traditional empirical model and an associated neural network. 
Thus, we coin our method, {\em neural material}.

Figure~\ref{fig:tease} shows a preview of our approach and results.
We demonstrate the performance of our approach on several problems, including coarsening applications, and synthetic examples that loosely resemble what could be available in various captured data scenarios.
% and captured data.  
We show how the neural network is trained and then reinforces the constitutive model to enhance the performance of the forward simulation.
%
%Furthermore, we show that by using well-structured training data (namely, simulated continuous deformation trajectories), the network can be trained with only a relatively small amount of training data. 
% PGK: Still not sure how much I believe this last sentence....
% PGK2: Given Uri's comments below... I'll take the liberty to suppress the last sentence.
%{\color{blue} Uri says: This introduction reads very well, and I like the new teaser! However, note there is no mention of coarse-fine FE meshes (which does appear in the teaser). I think that we can live with this, but just wanted to note it in this removable comment. Also, it is unclear that we want  to keep the last sentence above (about using few data), having no examples with real data..}

\section{Related work}
\label{sec:related}

\paragraph{Deformation Modeling}
Specifying material properties of a deformable object in order to yield a desired deformation behavior
is a common challenge in computer animation and physics-based simulation.
%It is a common challenge in computer animation and physics-based simulation to assign material parameters accurately in order to yield desired deformation behavior. 
Manual parameter tuning cannot scale to complex models with nonlinear or inhomogeneous material distributions. With recent improvements in sensing technologies, the data-driven approach of modeling and reconstructing deformation parameters from real world measurements has offered great potential for computer graphics applications, such as fabrics, soft objects, and human organs and faces~\cite{Pai01,Schoner04,Becker07,Wang11,Miguel12,Bickel09}. Bickel et al.~\shortcite{Bickel09} fit material parameters with an incremental loading strategy to better approximate nonlinear strain-stress relationships. Wang et al.~\shortcite{Wang11} proposed a piecewise linear elastic model to reproduce %the 
nonlinear, anisotropic stretching and bending of cloth. Miguel et al.~\shortcite{Miguel12} directly optimized %the 
nonlinear stress-strain curves based on measurements. Then \citet{Miguel2013} estimated internal friction.
Further, \citet{Miguel2016} developed a method for modeling example based materials with energy functions for both cloth and elastic solids.

A common weakness with previous methods is that they require a dense force displacement field.
% and known reference shapes. 
While Bhat et al.~\shortcite{Bhat03} avoided the need for force capture by using video tracking of cloth, they still assumed a trivial cloth reference shape. %By contrast, our method requires neither force displacement capture nor a priori reference shapes, making it significantly more convenient to use. 
{\color{black}Yang et al.~\shortcite{Yang2017} presented a learning-based algorithm to recover material properties of cloth from videos, using training datasets generated by physics simulators. 
However, 
\uri{their focus was still on material type estimation},
%the result is still only staying on material type estimation, 
due to inconsistency between real and synthetic data and sparse material space sampling. }
%Closely related to our work is that of~\citet{Wang2015}, which estimates 
% PGK: extra distance... given that we're not using surface scans???  Feel free to change back
\citet{Wang2015} estimated linear elastic material parameters from partially observed surface trajectories 
of an object's passive dynamics.  Our work has a similar setting, but focuses on correcting the errors that arise from assuming a linear elastic material and a simple damping force.

\paragraph{Material Design}
Material design has recently been gaining attention in the computer graphics community. In such design scenarios, physical objects are not always available for measurements, and specific and strict fabrication constraints must be respected.
Creating equivalent physics based models through numerical coarsening or model reduction can be seen as a related function approximation problem, which is less complicated than our problem given that we do not assume \uri{to} 
%that we 
have complete information.  \citet{Kharevych2009} took an energy based approach to coarsening composite elastic objects through the use of global harmonic displacements.
\citet{Nesme2009} created nonlinear shape functions and projected fine-level mass, stiffness, and damping matrices to produce coarse composite elements, 
while \citet{Torres2016} introduced an improved element based coarsening method that deals with co-rotation.  Coarsening techniques have proved useful for computational design for fabrication \cite{Chen2015,Panetta2015,Chen2017}, which must deal with the problem of modeling the behavior of real materials.
% PGK: might want to say something more about Chen 2017? 
In our work we were inspired by the design of nonlinear materials using principal stretches~\cite{Xu2015}.
{\color{black}One of the key ideas that enable simple design is formulating nonlinear material functions based on invariants of the deformation gradient, which leads to a simple and separable form of the energy equation. }
%This introduces the idea of having simple 
%energy 
%material functions based on invariants of the deformation gradient.
%and likewise takes care to ensure Hill-Drucker stability with a convex or polyconvex energy function.  
%One of the key ideas that enables simple design is the focus on isotropic materials, which leads to a simple and separable form of the energy equation.  
Our neural network corrections can be seen as fitting into a similar framework.
% mention damping design?? 
% or Interactive Material Design Using Model Reduction? ???

\begin{figure*}[htp]
	\includegraphics[trim=0 0.0cm 0 0.0cm,clip,width=\linewidth]{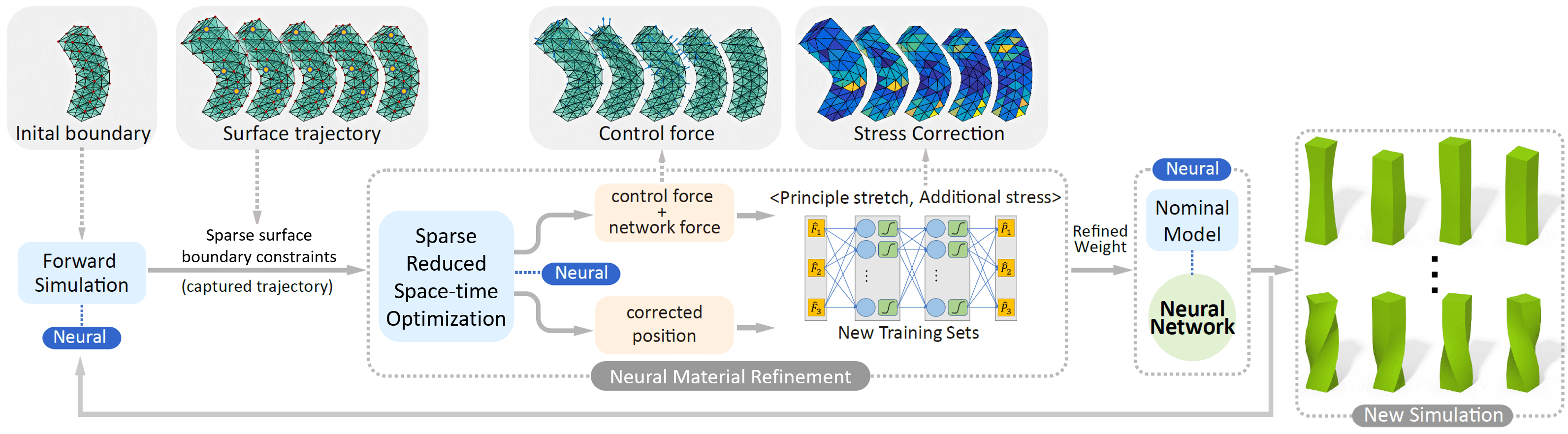}
	\caption{An overview of our process of learning a neural material model. Our neural material model learns a correction to a nominal material model that allows us to accurately reproduce the captured trajectory, even when the nominal model differs significantly.}
	\label{fig:loop}
\end{figure*}

\paragraph{Machine Learning in Material Science}
Neural networks have been previously investigated as a means %of
for computing complex stress-stress relationships of materials \cite{ghaboussi1991knowledge,Ghaboussi1998}. % PGK Wu 1995 thesis probably mostly covered by the earlier publication... 
\citet{Jung2006} modeled rate-dependent materials with neural networks, giving results both for a synthetic example and for data from a pre-stressed concrete beam.    
\citet{Stefanos15} used the length of strain trajectory traced by a material point, also called intrinsic time, as an additional input parameter in training. This is essential for situations of cyclic and transient loading. 

Capturing the elastic motion of objects can be time consuming.  In general, material modeling scenarios may not have many example trajectories to work from.  Additionally, the captured data is typically sparse because only the surface motion can be easily measured, and the visible surfaces will not typically cover the object.   
To overcome the challenge of training a convolutional neural network with a small dataset, \citet{Liang2017} employed a training strategy that combines three key ideas: unsupervised deep learning to determine the filter parameters of a convolution layer (generally using encoder-decoder based unsupervised learning strategies), supervised learning to determine the parameters in the classifier or regressor layer, and data augmentation to generate more training data.

Finally, we note that common simulation approaches, such as semi-implicit backward Euler that we use here, have artificial damping that depends on the time step size.  
% We will have damping that depends on the step size too... unless we swap out all our integrators *everywhere* :/
Recently, typical applications involving control and 3D printing have brought about the demand
for more quantitative simulation results, and methods involving little or no artificial damping have been employed~\cite{Chen2017,ChenAscherPai2017}.
%have become popular~ 
%\dc{if it is popular then I would expect more reference. How about simply ending the sentence after ``damping''?.}
In these methods, simulating the observed damping is largely delegated to the true damping force associated with the simulation.
% PGK: should come back to this, I don't think it is about the artificial damping..  

%:
\section{Overview}
\label{sec:overview}

%\begin{figure*}[htp]
%	%\includegraphics[clip,width=0.90\textwidth]{Pipeline.pdf}
%	\includegraphics[trim=0 0.0cm 0 0.0cm,clip,width=\linewidth]{Pipeline2.png}
%	%\includegraphics[trim=1.3cm 5.6cm 1.8cm 5.6cm,clip,width=\linewidth]{Pipeline.pdf}
%	\caption{An overview of our process of learning a neural material model. Our neural material model learns a correction to a nominal material model that allows us to accurately reproduce the captured trajectory, even when the nominal model differs significantly.}
%	\label{fig:loop}
%\end{figure*}

The core of our approach is to learn a function that corrects a nominal model of the elastic and damping properties of an object.  
Combining the learned function and the nominal model, {\color{black}the %so-called 
\uri{resulting} {\em{neural material}}} allows us to correctly reconstruct a sparsely specified desired trajectory, 
and compute new simulations that more accurately predict the true behavior of the 
%\dc{captured} 
deformable object.

We start with a tetrahedral mesh of the object for which we would like to learn accurate elastic and damping properties.  
For the nominal constitutive model, we use co-rotational linear elasticity and Rayleigh damping.  We first assign parameters to the nominal model. Values could be computed, for instance, with the methods of \citet{Wang2015}.  We use semi-implicit backward Euler 
integration with a fixed time step throughout our approach; thus, our simulations are stable but will suffer from numerical damping artifacts. 

%%The main loop \dc{why to refer to it as ``main loop here?} of our learning strategy 

For learning a material we %can 
use a set of incomplete surface trajectories of an object moving dynamically, unforced, in response to an initial perturbation. In this work, we use synthetic example data.  A typical synthetic capture sequence consists of immobilizing part of the object, while momentarily pushing another part of the object to form a static deformed state.  The captured trajectory is of the object as it returns to rest.  

The main loop alternates between solving a sparse reduced space-time optimization problem, and training of a neural network function (see Figure~\ref{fig:loop}).  
%The 
%\uri{A} key idea is that the sparse reduced space-time optimization uses a sparse selection of the incomplete surface trajectories to constrain nodes of the tetrahedral mesh, in addition to the normal physics constraints. 
% gentle control forces derived through space-time optimization can be 
%identify 
%\uri{identified} as what is currently missing in the system due to constitutive model inaccuracy. Then the neural network learns to distill the results of the space-time optimization and generalizes it by encapsulating it 
%into {\uri{neural material}}.
%The strain and strain-rate trajectory from the optimization, combined with the vertex 
%control forces, provide training data, but using this directly requires a complicated loss function.  We first solve a least squares problem to identify the stress on each tetrahedron from the vertex control force error. 
%As can be observed in the figure, the forward simulation and the space-time optimization are both assisted by the current {\uri{neural material}}. As the iterations progress, forward simulation can generate better initial seed trajectory and reduced modes to feed space time optimization; the neural material is retrained and reinforces with data of progressively better accuracy, which in turn improves the forward simulation.} 

%We compute an initial seed trajectory for this simulation using a forward simulation,

The key idea is that the neural network learns to distill the results of the space-time optimization and generalizes it by encapsulating it 
into neural material. As can be observed in the figure, 
the forward simulation and the space-time optimization are both assisted by the current neural material. As the iterations progress, the neural material is retrained and reinforces with data of progressively better accuracy, which in turn improves the forward simulation.
 %, and the space-time thereof.} % PGK cut this off early.. awkward?
 
 %%% PGK: THIS IS NOT A BAD FIGURE, BUT the other one was better here in the overview, so this one (not referenced in the paper, is now suppressed)
 %\begin{figure}
%	\includegraphics[width=\linewidth]{ConstitutiveModel.pdf}
%	\caption{Conventional nominal constitutive material model VS. Neural material model.}
%	\label{fig:NeuralMaterialModel}
%\end{figure}

% PGK: this might work nicely as an algorithm environment description... 

%\dc{now we are ready to go into more details...}

The sparse reduced space-time optimization uses a sparse selection of the incomplete surface trajectories to constrain nodes of the tetrahedral mesh, in addition to the normal physics constraints.  We compute an initial seed trajectory for this simulation using a forward simulation, which likewise has desired surface trajectories constrained to follow their known positions, in addition to the constraints on the immobilized parts of the mesh.  The starting pose of this forward simulation is computed as a static equilibrium using the constraints (immobilization and the set of incomplete example surface positions at time zero), the nominal elastic model, and the current neural network correction.  The static initial pose and constrained forward simulation give a good starting point for the sparse reduced space-time optimization, which quickly converges to a solution that identifies a plausible trajectory for unobserved nodes, and the corresponding gentle control forces. %\dc{gentle? micro?}

The gentle control forces identify what is currently missing in the neural network correction function.  
%\dc{I am not sure the reader understand by now that the neural network is merely a ``correcting'' function. It should have been said earlier... probably right where introduce ``Neural Material'' above; I think it is sort of explained next...}
% PGK: It looks like the text that was added above will perhaps address this concern??  
The strain and strain-rate trajectory from the optimization, combined with the vertex 
control forces provide training data, but using this directly requires a complicated loss function.  We first solve a least squares problem to identify the stress on each tetrahedron from the vertex control force error.  This is then added to the current neural network correction to produce training data samples.

We train a neural network to fit the strain, strain-rate, stress data.  Because the training data comes from gentle control forces, it may not be entirely self consistent (i.e., an element might need different forces to correct a given state of strain and strain-rate at different parts of the trajectory), but we let the network fit this data as best as possible; 
thus, we are learning an average correction. 
% fit the average response necessary to best reduce our loss function.  \dc{what is the loss function? was it described?} 
% PGK: perhaps removing the word loss makes this better?  
We have identified a few possibilities for the network architecture suitable for this problem, and largely use a two hidden layer one output layer network, with six nodes in each hidden layer.  
%All hidden nodes use a sigmoid activation function.  
% Seems they are ReLU, or something else now?  This can simply be suppressed.
We have explored training networks that output energy (i.e., the loss function compares the stress to the gradient of the network output), but using implicit integration with this architecture requires second derivatives of the network, which is costly, thus we learn a network that outputs stress directly.  
% PGK: do we need to note that this is dangerous

Once training is complete, we repeat the whole loop, starting by computing a forward simulation with {\color{black} updated {\uri{neural material}}}.
%the nominal material and newly trained neural network correction. 
We evaluate how well this trajectory matches the {\color{black} surface boundary constraint} 
%scan data
to determine convergence.  We can similarly monitor the magnitude of gentle control forces identified by the space-time optimization at each loop, and continue to iterate  as long as we see improvement at the space-time optimization step, even if the forward simulation error alone does not reveal that progress is being made.

\section{Neural Material Model}
\label{sec:alg}
{\color{black}
Standard constitutive material families such as 
%linear corotational, 
\uri{the} nonlinear St.Venant-Kirchhoff, Neo-Hookean, Ogden or Mooney-Rivlin materials 
%only cover small range of infinite-dimensional material spaces. 
\uri{do not account for all deformation phenomena that may arise.}
The main purpose of our proposed neural material 
%model 
\uri{method} is to use the function representation ability of neural networks to encapsulate variation of different materials in a unified way. 
 }
\subsection{Nominal Material Model and Assumptions}
{\color{black}
Our deformable models are 
%made of tetrahedra with 
\uri{constructed using} linear shape functions.
\uri{In order to handle large deformations of soft objects,}
% and 
the nominal material is \uri{described in terms of the} 
widely adopted co-rotated linear FEM,
% to handle large deformations of soft objects, while 
formulated using principal stretches \cite{Xu2015}.
}
%a linear elastic model formulated using principal stretches \cite{Xu2015}.
The ensuing computation of the element stresses and vertex forces is straightforward.  The deformation gradient $F$ for each tetrahedron is 
%then 
diagonalized by SVD, $F = U\hat{F}V^T$, and the Piola-Kirchoff stress is 
%first 
computed with the principal stretches, 
$\hat{P}(\hat{F})=2\mu(\hat{F}-I)+\lambda tr(\hat{F}-I)I$
where $\mu$ and $\lambda$ are Lamé parameters.  The diagonal stress is then transformed back to the world frame, $P=U\hat{P}(\hat{F})V^T$.  An element's contribution to its vertex forces is $P B_m$, where $B_m$ is the inverse material space shape matrix (see \citet{Sifakis2012}).  Summing the contribution of all elements, we can build a large sparse matrix $\mathsf{B}$ that combines the entries in $U$, $V$, and $B_m$ which can be multiplied by the block vector of all element diagonal stresses $\mathsf{\hat{p}}$ to give a block vector of all vertex forces $\mathsf{f}$, that is,  $\mathsf{B\hat{p}=f}$.

For implicit integration, we note that the gradient of stress $P$ with respect to the deformation gradient $F$ can be computed by the product rule and a careful evaluation of the different terms \cite{Xu2015}.  %In creating our synthetic test examples, we
%Our simulated examples likewise use the stiffness in the computation of Rayleigh damping, $(\alpha \mathsf{M} + \beta \mathsf{K})\mathsf{v}$.
%, with $\alpha = 0.01$ and $\beta=0.02$. % PGK <- Best to give the numbers in the results!!!
%
%\begin{equation}
%\frac{\partial P}{\partial F} = 
%\frac{\partial U}{\partial F} P(\hat{F}) V^T +
%U \frac{\partial{P(\hat{F})}}{\partial F} V^T +
%U P(\hat{F}) \frac{\partial{V^T}}{\partial F}.
%\end{equation}

%We further approximate the damping stress with the omission of the time derivatives of $U$ and $V$ assuming they are small. Thus, our nominal damping stress is computed as $U(2\beta\hat{\dot{F}}+\alpha tr(\hat{\dot{F}})I) V^T$.
% PGK: Irving comments about how this does not preserve angular momentum in Section 6.2
% PGK: Turns out we're not using this kind of damping at all, but the issue comes up 
% in another way as described below. 

\subsection{Network Based Material Model}
Following \citet{Irving2004}, damping force can also be implemented by transform the deformation gradient velocity $\dot{F}$ by the same $U$ and $V$ as used to diagonalize $F$, computing the damping stress $\hat{P}$ in rotated frame, and computing the force exactly as for preceding principle stretch-based elastic case.  
 
%Following \citet{Irving2004}, we transform the deformation gradient velocity $\dot{F}$ by the same $U$ and $V$ as $F$.  While the rotated velocity gradient is not generally diagonal, we set \smash{$\hat{\dot{F}}$} with only the diagonal terms of $U\dot{F}V^T$. 
%
Our network based material model computes correction to the Piola stress using the same rotations $U$ and $V$ as the nominal model, and corrects the elasticity and damping simultaneously. 
%In particular, note that we only provide principal stretches and the diagonal terms the diagonalized deformation gradient velocity \hat{\dot{F}}. 
%Our network based material model corrects the elasticity and damping simultaneously. 
Thus we define the function $N$ to be our neural network correction of the diagonal stress, \smash{$\Delta \hat{P} = N(\hat{F},\hat{\dot{F}})$}.
Using principal stretches and the diagonal deformation gradient velocity reduces the complexity of our function approximation problem.  As we will discussed later, 
the 6-input 3-output function approximation problem allows us to select a network architecture with a moderate number of hidden nodes, which can be trained with reasonable quantities of training data that we can easily produce.  
%PGK: might want to remove or revise this based on text in the NN sections??  Seems it is desirable to keep after a discussion with Bin.
%  We describe how the nerual network augments this material model, and how we compute the stiffness matrix needed in implicit integration. 

We do not currently include Rayleigh damping in our nominal model, thus the corrected stresses in the world frame are %computed as
\begin{equation}
P_n = U(\hat{P}(\hat{F}) + N(\hat{F},\hat{\dot{F}}))V^T,
\label{eq:force}
\end{equation}
and the gradient for implicit integration here includes a gradient of $N$, which is easily computed from the neural network function with automatic differentiation.

\section{Method}
\label{sec:alg}

%\textcolor{red}{We need a short introduction here, but likewise, perhaps there needs to be a section on the acquisition and preparation of data.}

In the following subsections, we provide in depth details on the different steps of our algorithm.  With many of the steps involving approximations, 
%and the captured surface trajectories may include noise.  As such, 
the general philosophy of the algorithm is to gradually learn the corrections necessary to produce forward simulations that replicate the 
%capture
desired trajectory. 

% next line is for debugging purposes.
%\newpage

\subsection{Sparse Reduced Space-Time Optimization}
% PGK : Removed constraints!  Watch out for this elsewhere! 

\newcommand{\norm}[1]{\left\lVert#1\right\rVert}
\newcommand{\bsfs}{\bm{\mathsf{s}}}
\newcommand{\bsfx}{\bm{\mathsf{x}}}
\newcommand{\bsfz}{\bm{\mathsf{z}}}
\newcommand{\bsfC}{\bm{\mathsf{C}}}
\newcommand{\bsfCf}{\bm{\mathsf{C_f}}}
\newcommand{\bsfCx}{\bm{\mathsf{C_x}}}
\newcommand{\bsfCz}{\bm{\mathsf{C_z}}}

The desired surface trajectory provides a rich source of information about the dynamics of the object.  The purpose of using a space-time optimization is to compute a set of gentle control forces that will correct our currently estimated neural material such that the simulation follows captured data.    
Many variations of the space-time constraints approach of \citet{Witkin1988} have been proposed.  To deal with the large number of degrees of freedom in our deformable models, we use reduction and sparse constraints taking inspiration from \citet{Barbic2009} and \citet{Schulz2014}.

We compute a reduced basis $\Phi$ from principal component analysis (PCA) of a trajectory created with an unconstrained forward simulation with the provided initial conditions and our current approximation of the material model.  We desire a small yet expressive basis which is fast to compute.  For the analysis we use a short portion of the forward simulation sequence which targets the most interesting dynamics, and perform PCA on only a fraction of the frames, in turn keeping only a fraction of the vectors for the basis.  For instance, in many examples we use one fifth of the frames of approximately a half second of simulation at 1000 Hz, keeping half of the vectors for a 50 dimensional reduced basis.   

Our objective function consists of two parts: physical constraints, and sparse trajectory constraints.
%$\bsfC$ contains both physics constraints $\bsfCf$ and position constraints $\bsfCx$. 
Using a discretized approximation of the acceleration, the unreduced equation of motion at time step $i$ is given by
\begin{equation}
%\mathsf{C}_i \equiv 
h^{-2}\mathsf{M}(\mathsf{x}_{i-1} - 2\mathsf{x}_i + \mathsf{x}_{i+1} ) 
= \mathsf{B}_{i+1} \mathsf{p}_{\mathsf{n},i+1} 
+ \mathsf{f}_{ext},
%-\mathsf{f}_{i+1}.
%%%\hat{P}(\hat{F_j}) + N(\hat{\dot{F_j}},\hat{\dot{F_j}}) ) 
\label{eq:physicsconstraint}
\end{equation}
with gravity force $\mathsf{f}_{ext}$.  This equation corresponds to our %semi-implicit backward Euler 
forward integration method because the force term evaluation is at the
end of the time step.  However, we optimize with reduced coordinates $\mathsf{z}_i$, where $\mathsf{x}_i = \Phi \mathsf{z}_i$, thus, the reduced physics constraints are
\begin{equation}
\mathsf{C_f}_i \equiv 
h^{-2}\Phi^T\!\!\mathsf{M}\Phi(\mathsf{z}_{i-1} - 2\mathsf{z}_i + \mathsf{z}_{i+1} ) 
- \Phi^T\mathsf{B}_{i+1} \mathsf{p}_{\mathsf{n},i+1} 
- \Phi^T\mathsf{f}_{ext}.
\label{eq:Cfi}
\end{equation}
While desired example trajectory may already be sparse, such as an incomplete scan of the surface, we ultimately only need an extremely sparse set of position constraints.  The sparse position constraints we use are defined with a random but well distributed set of points from the desired trajectory (we typically use 5 or 6 points).  Letting vector $\mathsf{s}_{i}$ contain the desired point positions at time step $i$, we can write the sparse trajectory constraints as
\begin{equation}
\mathsf{C_z}_i \equiv \lambda( \mathsf{S} \Phi \mathsf{z}_i - \mathsf{s}_{i} ),
\end{equation}
where the wide sparse selection matrix $\mathsf{S}$ extracts the components of the desired positions by having one non-zero entry per row.  The scalar $\lambda$ is used to specify the weight of position constraints given that the combination of physics and position constraints are solved in a soft manner.

Letting $\bsfC$ concatenate all physics constraints $\bsfCf$ on top of all position constraints $\bsfCz$, our goal is to find a reduced trajectory $\bsfz$ that minimizes the violation of all constraints.  We approach this as a root finding problem, with the starting point being the projection of the constrained forward simulation trajectory into the reduced basis.  Thus, we iterate on solving     
%In our formulation, by choosing vertex control forces instead of stresses, $\hat{\bsfs}$ will always be zero.  Furthermore, we simplify the solution of Equation~\ref{eq:ST2} to 
\begin{equation}
\frac{\partial \bsfC}{\partial \bm{\mathsf{z}}} \Delta {\bm{\mathsf{z}}} = -\bsfC, 
\label{eq:ourSTC}
\end{equation}
updating our solution $\bm{\mathsf{z}}^* \leftarrow \bm{\mathsf{z}}^* + \zeta_\mathsf{z} \Delta \bm{\mathsf{z}}$, using a step damped by factor $\zeta_\mathsf{z}$, typically in the range 0.1 to 0.5.
%, 
%and then directly setting all constrained positions $\bm{\mathsf{z}}_\mathsf{c}$ in $\bm{\mathsf{z}}^*$ to satisfy
%$\bsfCx$.

We do not assemble the gradient matrix $\bsfC$ in Equation~\ref{eq:ourSTC}, but use the chain rule and keep it in the factored form,
%\smash{\frac{\partial \bsfC}{\partial \bm{\mathsf{z}}} = 
$\smash{\frac{\partial \bsfC}{\partial \bm{\mathsf{x}}} \frac{\partial \bsfx}{\partial \bsfz},}$
%\begin{equation}
%\frac{\partial \bsfC}{\partial \bm{\mathsf{z}}} = \frac{\partial \bsfC}{\partial \bm{\mathsf{x}}} \frac{\partial %\bsfx}{\partial \bsfz},
%\end{equation}
where $\smash{\frac{\partial \bsfx}{\partial \bsfz}}$ simply contains copies of the basis matrix $\Phi$.
% since $\frac{\partial \mathsf{x}_i}{\partial \mathsf{z}_i}=\Phi$.
%The gradient $\smash{\frac{\partial \bsfCf}{\partial \bm{\mathsf{x}}}}$, which is 
The top part of gradient $\frac{\partial \bsfC}{\partial \bm{\mathsf{x}}}$, that is, the gradient of $\bsfCf$, has a very simple part that links vertices at different time steps through the acceleration term, and a more complex part where the chain rule must be applied to compute the force gradient, which includes a contribution from the neural network.  This second part sprinkles off diagonal terms into the matrix linking vertices that are adjacent to a common element.  The bottom part of gradient $\frac{\partial \bsfC}{\partial \bm{\mathsf{x}}}$, that is, the gradient of $\bsfCz$, simply contains copies of the selection matrix $\mathsf{S}$.  The resulting matrix is tall, and while the constraint gradient matrix is very large, it is also very sparse, and we compute the solution using the Eigen library's sparse least squares conjugate gradient implementation.
% PGK: CAN'T be true... it is not square... it is a tall matrix and needs a least squares appraoch.
\begin{figure*}[t!] % trim left lower right uper
	\includegraphics[trim=0 1.0cm 0 0.5cm,clip,width=\linewidth]{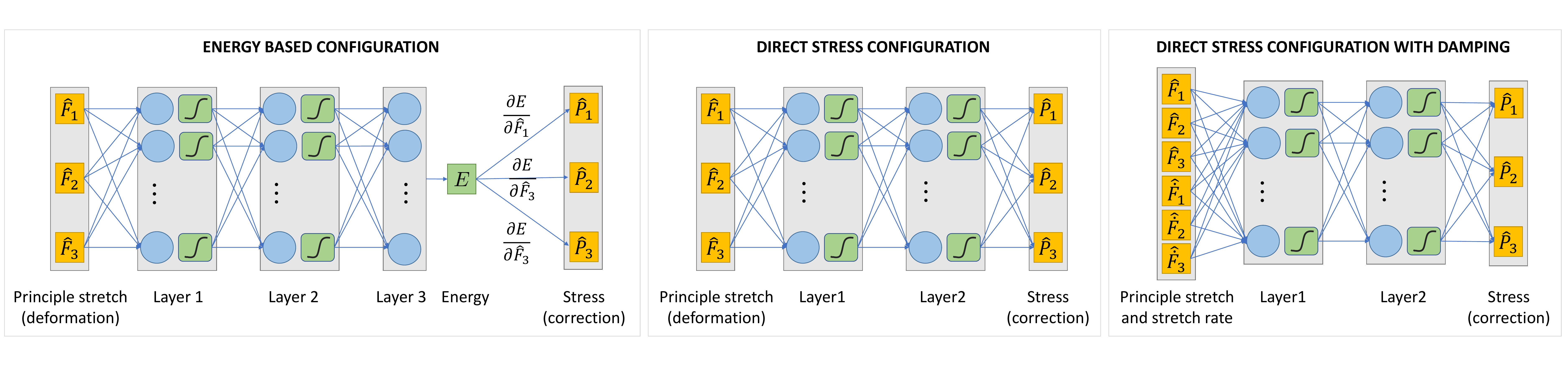}
	\caption{Example configurations for the networks used in our neural material model.  
		While the energy based configuration may have nice properties, we can still
		produce high quality corrections with a direct computation of stresses,
		which has the advantage of a much less expensive material stiffness computation.
		Our networks have 6 neurons in the hidden layer when estimating
		an elastic correction alone, %6 inputs (both $\hat{F}$ and $\dot{\hat{F}}$) 
		and 9 neurons in each hidden layer
		% later 
		for a correction that includes damping.}	
	\label{fig:networkconfig}
\end{figure*}

We can check for convergence of solution $\bm{\mathsf{z}}^*$ by monitoring our progress in reducing the violation of physics constraints in Equation~\ref{eq:Cfi}.  Once converged, it is exactly these violations that provide the necessary control forces to correct our current neural material model.    
%While we do not even directly solve for the control force (our optimization only solves for the reduced coordinate trajectory), 
Given an optimized reduced trajectory $\bsfz$, the gentle control force is computed from the error in the physics constraints,
\begin{equation}
\mathsf{f}_{i+1} =  
h^{-2}\mathsf{M}\Phi(\mathsf{z}_{i-1} - 2\mathsf{z}_i + \mathsf{z}_{i+1} ) 
- \mathsf{B}_{i+1} \mathsf{p}_{\mathsf{n},i+1} 
- \mathsf{f}_{ext}.
%%%\hat{P}(\hat{F_j}) + N(\hat{\dot{F_j}},\hat{\dot{F_j}}) ) 
\label{eq:physicsconstraint}
\end{equation}
While it may be desirable to solve for control stresses at each element, as these are what are required for training, our approach permits an easier solution that directly provides a control force at each vertex.

\subsection{Training Data Preparation}

%%% PGK: THIS MAY SEEM EARLY IN THE .tex BUT IT MAKES IT APPEAR ON PAGE 5

The result of our space-time constraints optimization 
%(and likewise the constrained forward simulation in some cases), 
provides vertex control forces to correct our model, but we need stresses to train our neural network.
 
% Easy fix
At every time step $i$, we must identify stresses that produce the desired control forces $\mathsf{f}_i$.  While the linear system $\mathsf{B\hat{p}=f}$ has a tall matrix for a single tetrahedron, typical larger systems will have more tetrahedra than vertices and $\mathsf{B}$ will be a fat matrix.  Indeed it is possible for different combinations of stresses to produce the same force.  However, $\mathsf{B}$ also has deficient row rank as there can be forces that cannot be realized by internal stresses alone.
 (For instance, a constant force on all vertices as would be produced by gravity cannot be produced by internal stresses.)  
 Thus, we compute the stresses as a least squares problem using LSQR \cite{Paige1982}, which finds $\mathsf{\hat{p}^*}$ that minimizes  $\norm{\mathsf{\hat{p}}}$ subject to 
$\mathsf{B^TB \hat{p} = B^Tf}$ using an iterative algorithm that exploits sparsity and avoids assembling the product $\mathsf{B^TB}$.

During the training process, the neural network will have developed an estimate for the required correction to the diagonal Piola stress $\Delta \mathsf{\hat{P}}$.  Thus we combine the current network output with our newly estimated stress corrections.  For each time step, we solve the least squares problem, and then for each element extract its stress $\hat{P}^*$ from the block vector $\mathsf{\hat{p}}^*$, from which we then assemble a training data pair
\begin{equation}
\left(\hat{F}, \hat{\dot{F}}\right), \ \
\left(N(\hat{F},\hat{\dot{F}})+\zeta_P\hat{P}^* \right),
\end{equation}
where the factor $\zeta_P$, typically set to $0.1$, allows us to take smaller conservative steps towards learning the correction.

\subsection{Constrained Forward Simulation}

We use semi-implicit backward Euler integration for simulation.  While this choice of integrator has the disadvantage of time step dependent numerical damping, it is convenient due to the ease of implementation and stability.  At each step we solve the equation 
\begin{equation}
\mathsf{A} \Delta \mathsf{v} = h \mathsf{f}
\end{equation}
where $\mathsf{f} = \mathsf{B p_n} + \mathsf{f}_\text{ext}$, with $\mathsf{p_n}$ being the block vector of neural material stresses at the current time step, $\mathsf{f}_\text{ext}$ being the external gravity force, and $\mathsf{A = M - hD -h^2K}$, where $\mathsf{D}$ and $\mathsf{K}$ are assembled using the gradient of Equation \ref{eq:force}.  Many of our models are rigidly attached to the world, and we typically remove these degrees of freedom from the system.  Forward simulation is an important step in our fitting process as we use it to evaluate the performance of the current neural material estimate, which we do by monitoring the maximum vertex error compared to the ground truth trajectory observation.

% PGK unsure what to call this at this point... 
The sparse reduced space-time optimization needs a reasonable starting trajectory.  While the forward simulation with the current neural material estimate could serve this purpose, we find it 
%it is more 
valuable to simulate a trajectory constrained to follow the desired surface motion.  We can divide the vertices into two groups, 
\begin{equation}
\begin{pmatrix}
\mathsf{A_{uu}} & \mathsf{A_{uc}} \\ \mathsf{A_{cu}} & \mathsf{A_{cc}}
\end{pmatrix}
\begin{pmatrix}
\Delta \mathsf{v_u} \\ \Delta \mathsf{v_c}
\end{pmatrix}
=
h
\begin{pmatrix}
\mathsf{f_u} \\ \mathsf{f_c}
\end{pmatrix}
\end{equation}
where we use subscript $u$ for unconstrained and $c$ for constrained.  The second block row can be discarded leaving us a smaller system to solve, namely,
\begin{equation}
\mathsf{A_{uu}} \Delta \mathsf{v_u} =  h\mathsf{f_u} - \mathsf{A_{uc}} \Delta \mathsf{v_c}.
\end{equation}
Here, $\Delta \mathsf{v_c}$ at time step $i$ is computed by a second order central finite difference, $h^{-1}(x_{i-1} - 2x_i + x_{i+1})$.  We solve these large sparse systems using PARDISO~\cite{pardiso-6.0a,pardiso-6.0b}.

%\begin{figure*}[t!] % trim left lower right uper
%	\includegraphics[trim=0 1.0cm 0 0.5cm,clip,width=\linewidth]{NetworkConfigurationsV3.pdf}
%	\caption{Example configurations for the networks used in our neural material model.  
%		While the energy based configuration may have nice properties, we can still
%		produce high quality corrections with a direct computation of stresses,
%		which has the advantage of a much less expensive material stiffness computation.
%		Our networks have 6 neurons in the hidden layer when estimating
%		an elastic correction alone, %6 inputs (both $\hat{F}$ and $\dot{\hat{F}}$) 
%		and 9 neurons in each hidden layer
%		% later 
%		for a correction that includes damping.}	
%	\label{fig:networkconfig}
%\end{figure*}

\section{Neural Network Design and Training}
\label{sec:net}

For every time %step
%level 
step
% PGK: why is this called a level?
and every 
finite element we have a data point with which to train the network.  
For small models and short sequences of captured motion, this can be on the order of thousands of points.  
There are two important and related questions: (i) how much data is necessary to train the network, and (ii) how many neurons and in what configuration do we need to successfully fit the data.

%\begin{figure*}[t!] % trim left lower right uper
%	\includegraphics[trim=0 1.0cm 0 0.5cm,clip,width=\linewidth]{NetworkConfigurationsV3.pdf}
%	\caption{Example configurations for the networks used in our neural material model.  
%		While the energy based configuration may have nice properties, we can still
%		produce high quality corrections with a direct computation of stresses,
%		which has the advantage of a much less expensive material stiffness computation.
%		Our networks have 6 neurons in the hidden layer when estimating
%		an elastic correction alone, %6 inputs (both $\hat{F}$ and $\dot{\hat{F}}$) 
%		and 9 neurons in each hidden layer
%		% later 
%		for a correction that includes damping.}	
%	\label{fig:networkconfig}
%\end{figure*}

\subsection{Network Design}
\label{sec:4.1}

We investigated the network configurations shown in Figure~\ref{fig:networkconfig}. 
The energy based neural network shown at left has the benefit of ensuring a conservative correction to the material (see \cite{Miguel2016}), but we found that the computation of the energy network Hessian, 
%{Jacobian}, 
% PGK: we need the Jacobian of the forces, or the Hessian of the energy.  Not clear??  :/  added the word energy
as needed for implicit 
time integration, is undesirably costly (even with the automatic differentiation methods in the latest version of PyTorch).  As such, we use the configuration shown at middle and right in the figure, which still ensures that element forces will not violate linear momentum conservation. 

\setlength{\columnsep}{9pt}
\begin{wrapfigure}[14]{r}{0.45\linewidth}%
	%\vspace{-2.05em}  %kinda a gross way to make it fit.. unnecessary if the following paragraph is not wrapped in 
	% curly braces as wrap fig will continue to insert properly.
	\centering
	\includegraphics[trim=1.2cm 5.6cm 2cm 5.6cm,clip,width=\linewidth]{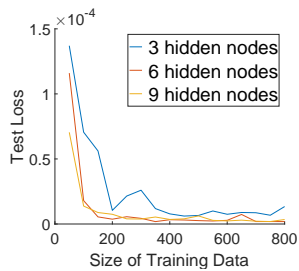}
	\caption{Evaluation of loss by number of hidden nodes.}
	\label{fig:loss_by_node}
\end{wrapfigure}%
We tested 3, 6, and 9 neurons in each hidden layer and trained the network to compute the stress of material models for a single element 
(co-rotational, Saint Venant Kirchhoff (StVK), neo-Hookean).
Note that the function we train here is not a correction, but the stretch to diagonal stress relationship 
(thus, we do not expect the loss function to completely vanish, because we are not using all six dimensions of the strain).  We use training data produced via random deformations generated by moving the vertices by a random displacement drawn from a Gaussian with standard deviation equal to 66\% the element size.  As described in Figure~\ref{fig:loss_by_node}, we observed 
that 6 neurons in the hidden layers performed better than 3, while increasing to 9 neurons did not show significant additional improvement.

\begin{figure}
	\includegraphics[trim=3.3cm 8.0cm 3.8cm 8.6cm,clip,width=.48\linewidth]{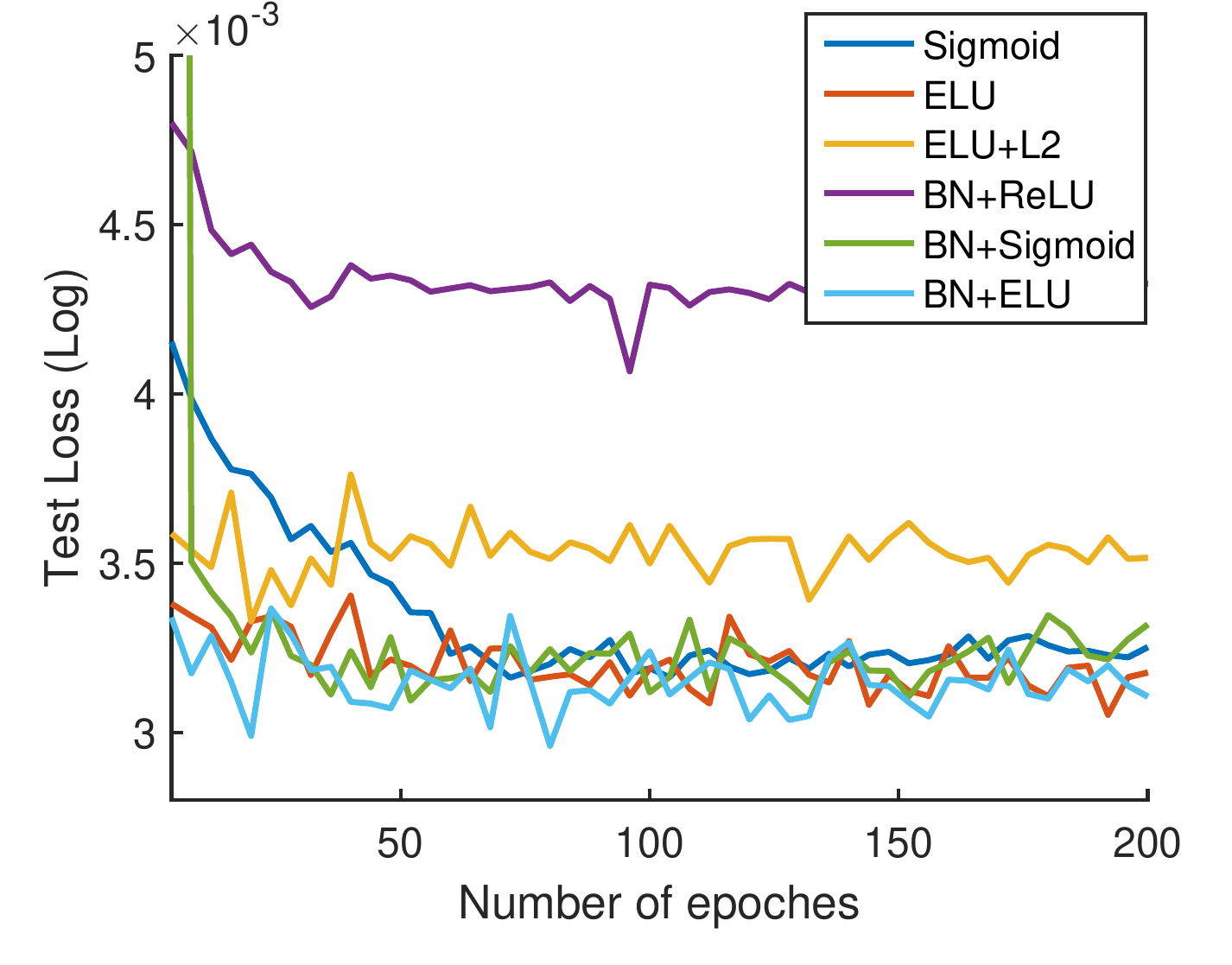}
	\includegraphics[trim=3.3cm 8.0cm 3.8cm 8.6cm,clip,width=.48\linewidth]{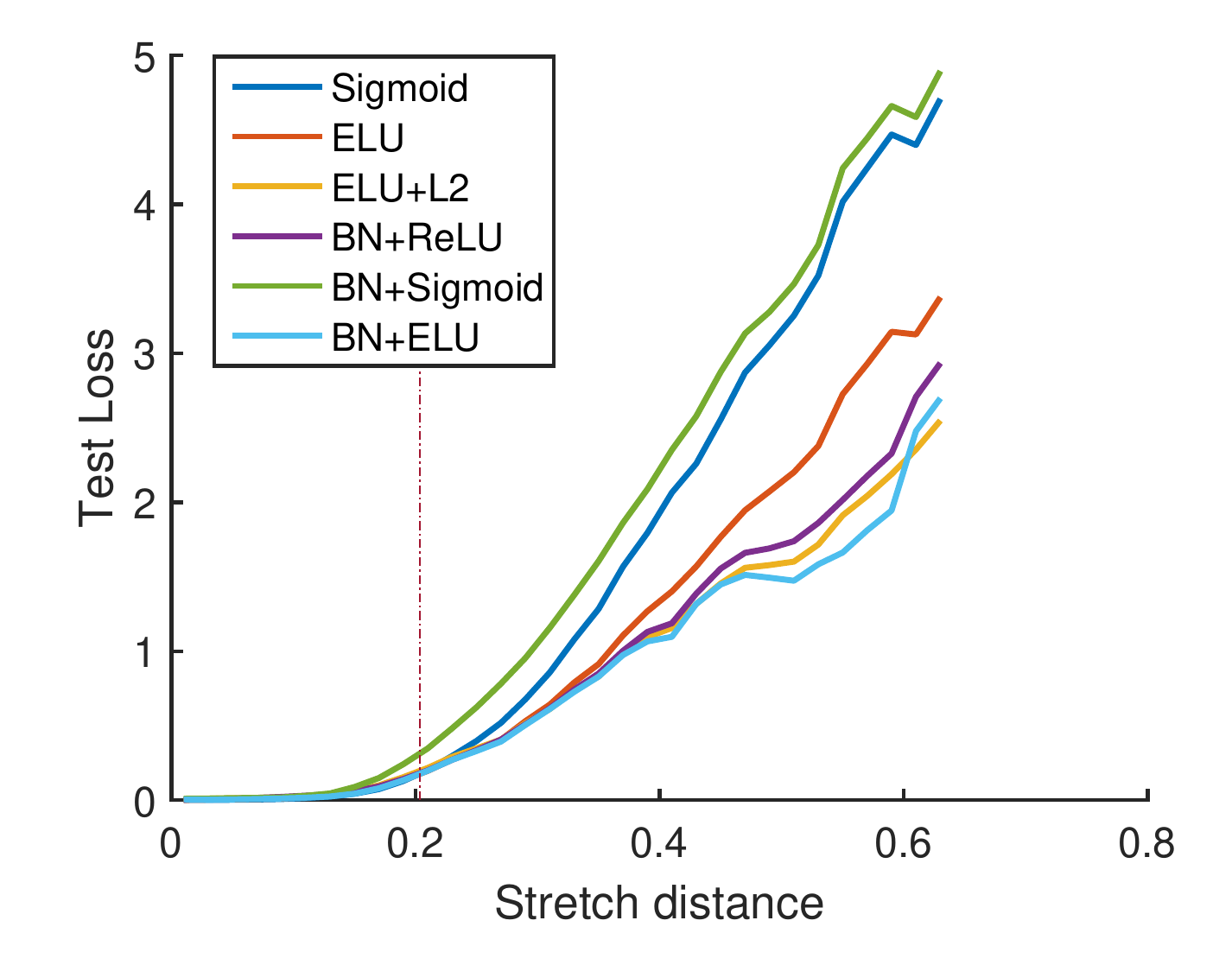}
	\caption{ Evaluation of activation functions in a neural network. Left: Network's learning rate and test accuracy with different active function configurations. 
	Right: Network's test accuracy along deformation scale of test data. We found that {the combination of} Batch normalization layer + ELU has 
	{superior} performance and robustness.}
	\label{fig:activation_result}
\end{figure}

%\color{red}
Having tested different activation functions with and without batch normalization,
we have settled on using 
	%\color{red} 
	ELU activation functions \cite{Clevert2015} after a batch normalization layer \cite{Ioffe2015} for better network performance and more robustness to noise. The evaluation can be seen in Figure~\ref{fig:activation_result}, 
where we used the same training data for all the network configurations, 
and the training data are generated from the first iteration output of 
	%\color{blue}
	the turtle example.  
As the left image of Figure~\ref{fig:activation_result} shows, most of the activation function configurations, except for BN+ReLu, exhibit similar performance on learning speed and accuracy when the test data has similar deformation scale as the training data. Here we use the distance between principal stretch $\mathsf{\hat{F}}$ and non-deformed principal stretch $(1,1,1)$ to evaluate the deformation scale as $\| \mathsf{\hat{F}} - (1,1,1) \|$. 
To show the expandability of a network, we tested the network with much larger scale deformation data. 
As demonstrated in the right image of Figure~\ref{fig:activation_result}, training data reside in the left side of red dot line, in the range of $[0,0.26]$. 
Beyond this range, ELU performs better than the sigmoid function.  
%\setlength{\belowdisplayskip}{1pt}

% wrapfig?
%\begin{figure} % trim left lower right uper
%\includegraphics[trim=3.5cm 8cm 4cm 8cm,clip,width=.8\linewidth]{loss_by_hidden_node.pdf}
%\caption{Evaluation of loss by hidden nodes. The improvement from 3 to 6 neurons is significant, but that from 6 to 9 is not.}
%	\label{fig:loss_by_node}
%\end{figure}

\subsection{Network Training}
\begin{figure}[h] % trim left lower right uper
	\includegraphics[trim=1.3cm 5.6cm 1.8cm 5.6cm,clip,width=.48\linewidth]{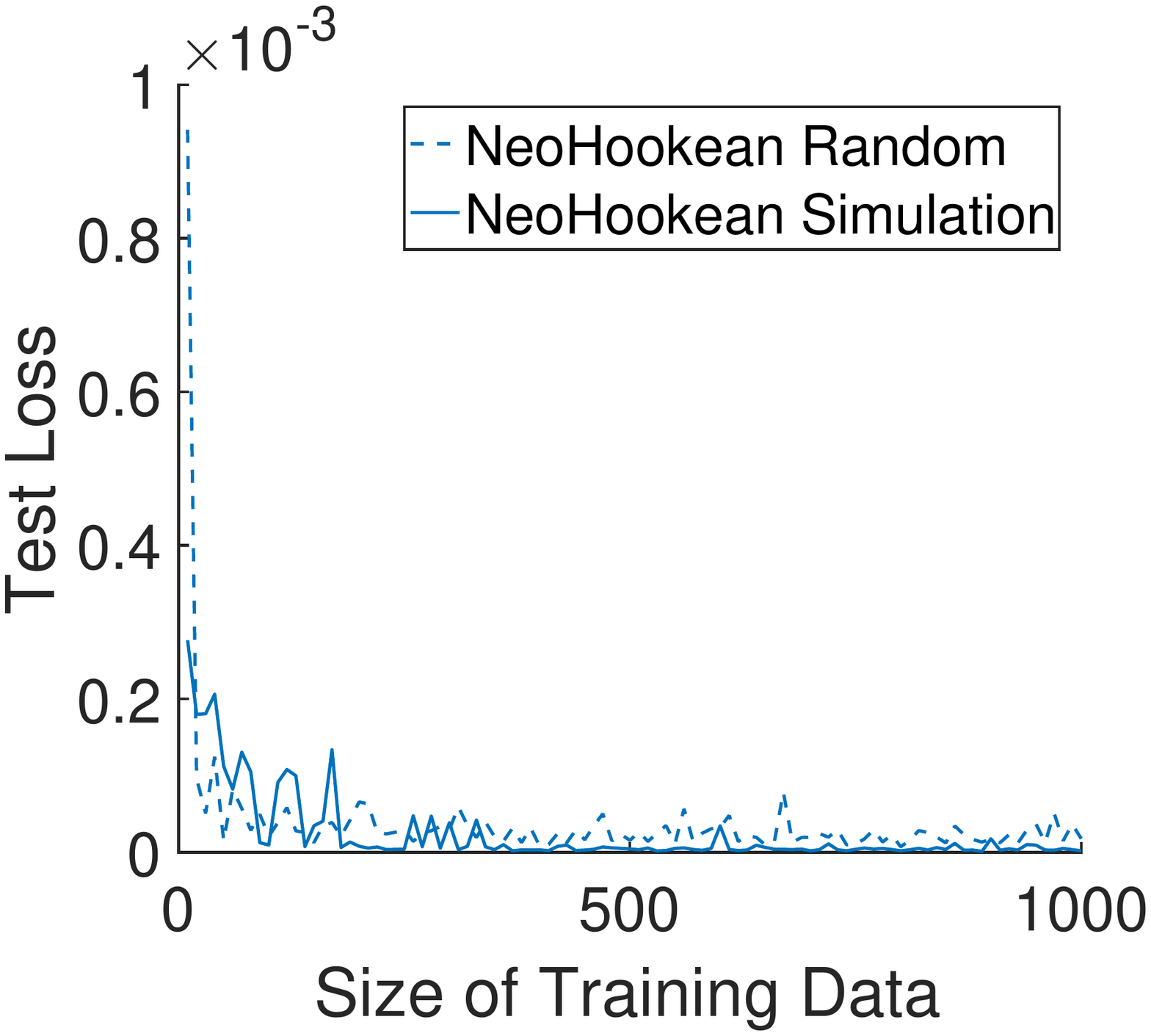}
	\includegraphics[trim=1.3cm 5.6cm 1.8cm 5.6cm,clip,width=.48\linewidth]{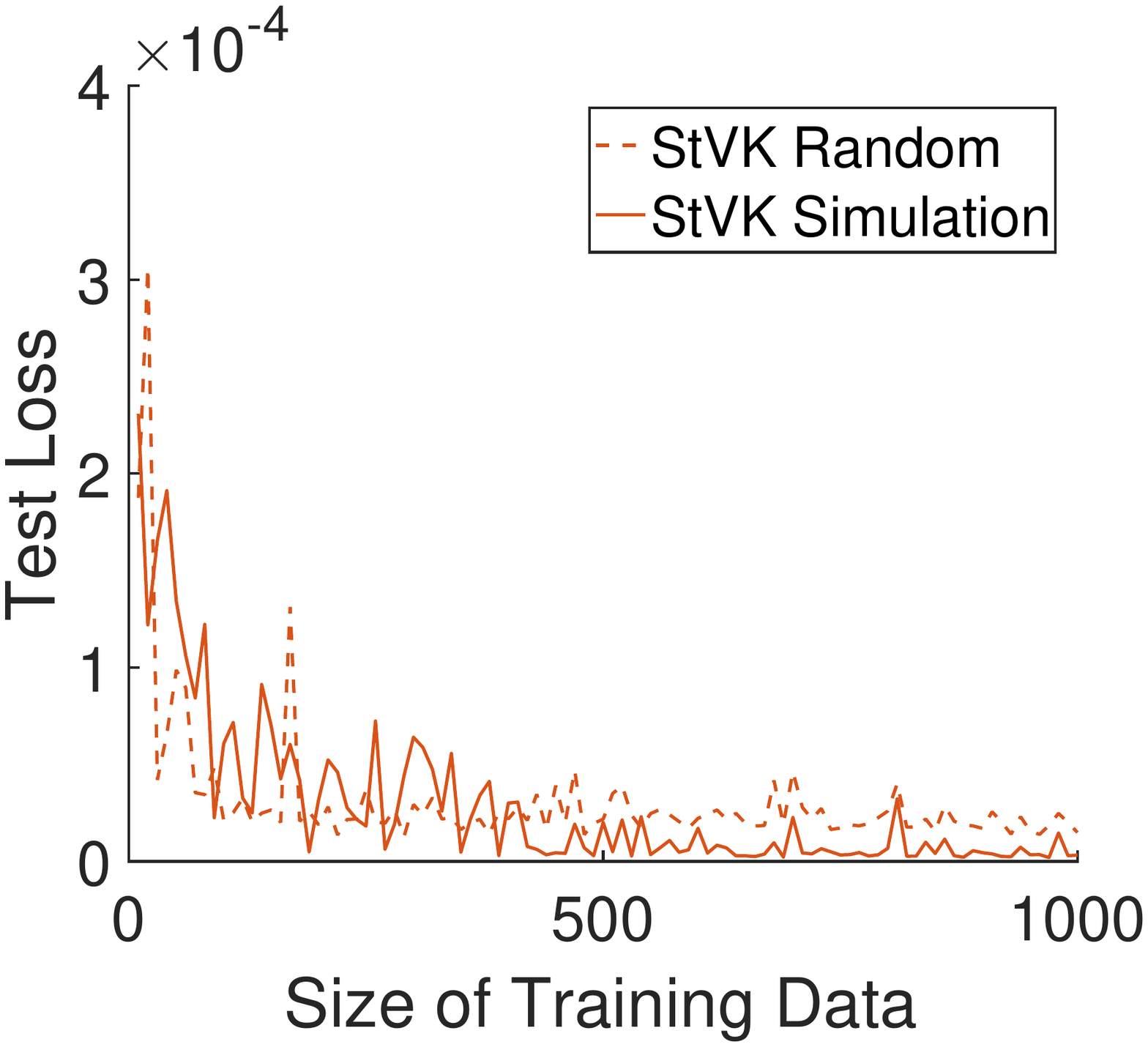}
	\caption{Evaluation of neural network learning for estimating Neo-Hookean and Saint Venant Kirchoff models using different amounts of training data.}
	\label{fig:stvkneohook}
\end{figure}

We follow standard practices in training our networks, computing scaling factors for the inputs and outputs based on the training data so that both inputs and outputs have zero mean and unit variance.  
%(\pk{40\% of the training data samples are held out for testing}).
We randomly permute the order of the samples across time and tetrahedra to improve training, and the network is retrained from scratch with each new collection of training data produced with sparse reduced space-time optimization.
%at each time step.  
% PGK: \pk{QUESTION: is there a good reason for or against not starting with the weights that we already had?}
% PGK ANSWER: Yuamin and Bin tried it and it got stuck in the weights that were given... just supressing the detail
% as I think it just brings up more questions (i.e., why)
When training networks to estimate standard nonlinear material models as listed 
%above, % PGK seems a bit far... not just above, but on the previous page! :/)
in Section~\ref{sec:4.1},
%(corotational, Saint Venant Kirchoff, Neo-Hookean), 
we find that a few hundred samples is sufficient to train these functions.  Figure~\ref{fig:stvkneohook} shows that this is the case for Neo-Hookean and Saint Venant-Kirchhoff materials.  The plots show the average performance across 3 trainings where the loss function is the norm of the difference of the network output across the test data set.  The training data either comes from random deformations, or from a simulated trajectory where an elastic bar is pulled away from its rest configuration.
The test data set in both cases is data from a new simulation sequence that was not seen in the training data.
These figures demonstrate that it is possible to learn these material models from a simulation sequence using a similar number of samples and achieve a result of similar quality to using random deformations.
%)similar quality result. \pk{ ARE THESE TEST LOSSES ON RANDOM OR SIMULATED DATA?  RANDOM MIGHT BE HIGHER?} 
%Here we could describe the tests that show that only a few hundred samples of training data are necessary 
While these learned material functions will not extrapolate to large strain, they still perform well in a region that involves significant deformation.  
Furthermore, these tests show success using training data that comes from simulation sequences, which suggests that we can likewise successfully capture constitutive models of unknown materials from captured sequences. 

%%%%%%%%%%%%%%%%%%%%%%%%%%%%%%%%%%%%%%%%%%%%%%%%%%%%%%%%%%%%%%%%
%%
%% PGK: Seemed a bit controvercial to pull this figure and text, but Bin suggested that extrapolation is covered in the video
%% and elsewhere in the paper (wrt to the video)... I'm not sure where... but I've put this all in comments now...
%%
%%
%\begin{figure}[h]
%	\centering
%	%\includegraphics[trim=1cm 5.6cm 2cm 6cm,clip,width=.485\linewidth]{trainingset.pdf}
%	\includegraphics[trim=0cm 0cm 0cm 0cm,clip,width=.75\linewidth]{netoutput.png}
%	\caption{Evaluation of the reliability of neural network learning from sparse simulation training data. 
%		{\color{blue} Displayed are the} distribution of training data (red start) and random test data (dots); the color on dots represents the scale stress error between neural network output and ground truth.  }
%	\label{fig:networkoutputerror}
%\end{figure}
%%
%Figure \ref{fig:networkoutputerror} further shows a comparison of the error distribution across random test data for a neural material trained from a simulated trajectory.   
%{\color{blue} Bin: I feel maybe this part is not needed any more. The extropolation of training result now is demenstrated in video with two different test trajatories. } 
%{\color{blue} Uri: Do we have active reason to change this at this late stage?}
%%
%%%%%%%%%%%%%%%%%%%%%%%%%%%%%%%%%%%%%%%%%%%%%%%%%%%%%%%%%%%%%%%%

\begin{figure}[h]
	\centering
	\includegraphics[width=.9\linewidth]{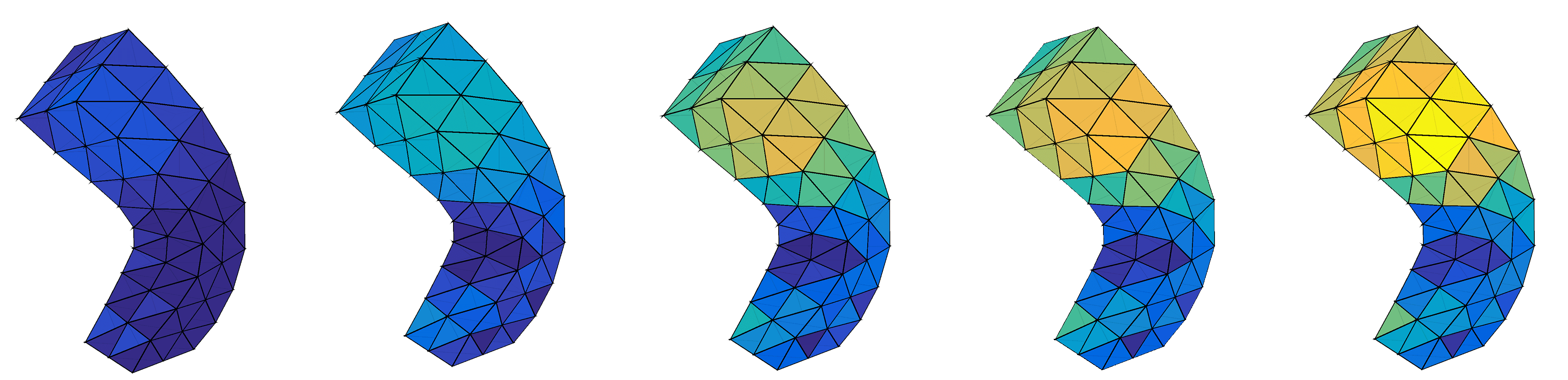}\\
	\includegraphics[width=.9\linewidth]{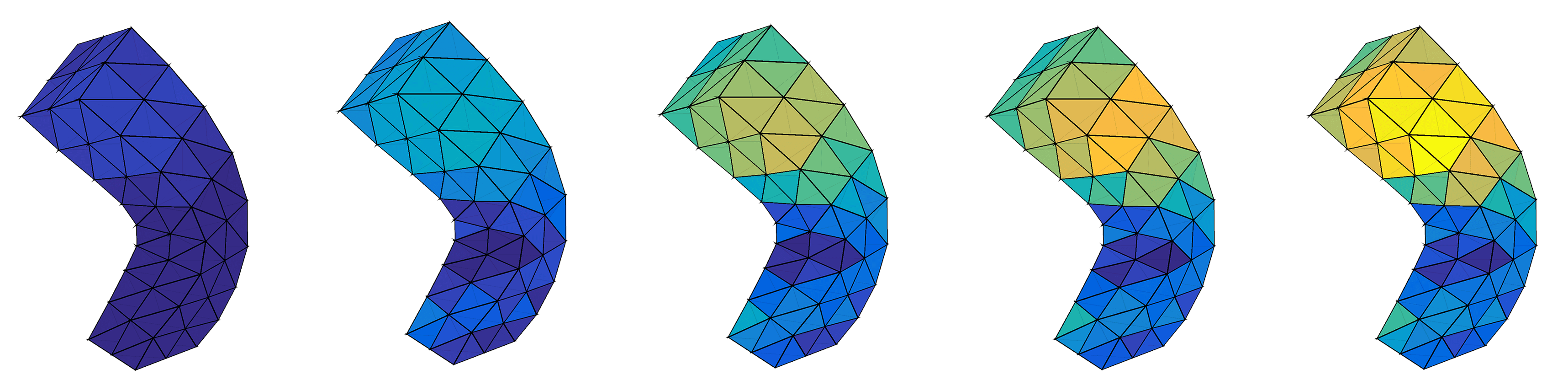}\\
	\hspace{1cm} 1 \hfill 5\hfill 9 \hfill 12 \hfill 15 \hspace{.75cm}
	\caption{The neural network progressively learns the necessary correction, seen here
		at different iteration numbers of our algorithm.   The top row shows stresses from space-time optimization for one example
		frame in the sequence, while the bottom row shows the corresponding learned corrections.  
		We observe that the neural network models important corrections, while 
		%likewise
		distilling conflicting training samples or noise that might be present in the training data.
	}
	\label{fig:prepost}
\end{figure}

Figure~\ref{fig:prepost} demonstrates that the neural network learns important corrections over the course of multiple iterations of our algorithm.  Furthermore, the neural network is able to distill any conflicting training data to provide a consistent correction.

%Discussion on the comparison simulation data in comparison to random samples in training, and generally questioning if the capture data provides enough information to learn the true material. 

%Ultimately a simple network is sufficient!

\section{Results and Discussion}
\label{sec:results}

In the following sections, we describe experiments that help reveal what is taking place in each step of the algorithm. 
%Finally 
We show results of our method in action and discuss a collection of material estimation scenarios. To validate the accuracy of our material model estimation algorithm, we use synthetic data generated by forward simulations with known elasticity parameters and damping properties.

%With so many steps in our algorithm, it is important to evaluate the performance of each.  In the sections below, we describe experiments that help reveal what is taking place, as well as other experiments that allow us to make choices for the neural network design.  Finally we show results and discuss a collection of material estimation scenarios.

\subsection{Space-Time Optimization}
\begin{figure}
	\includegraphics[trim=0cm 0cm 0cm 0cm,clip,width=.4\linewidth]{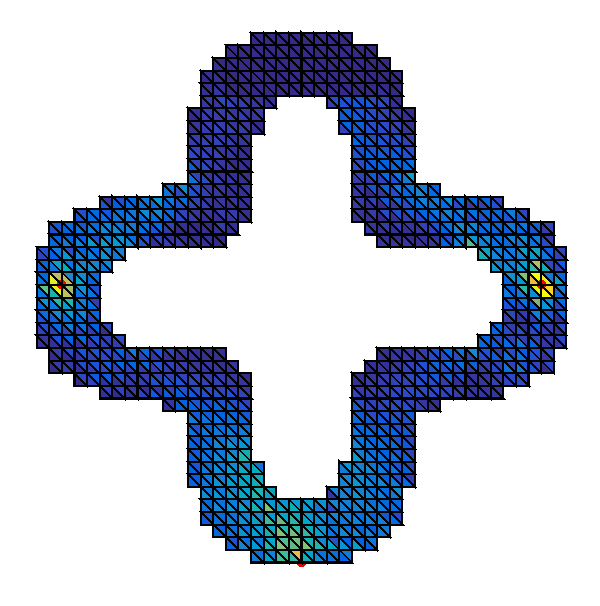}
	\includegraphics[trim=0cm 0cm 0cm 0cm,clip,width=.4\linewidth]{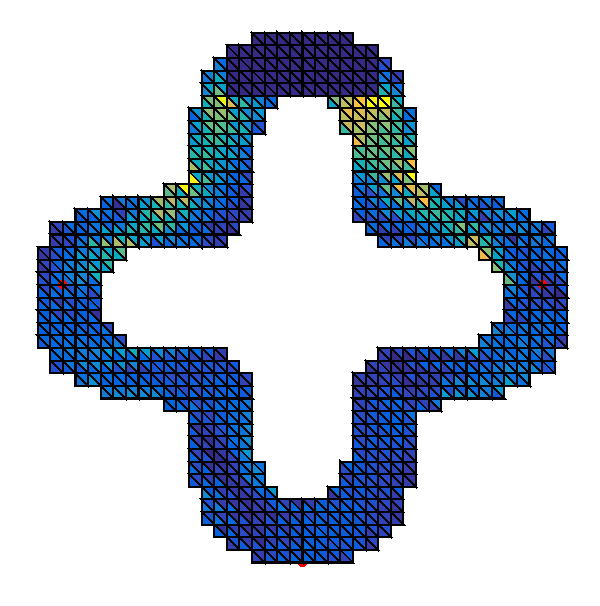}
	\includegraphics[trim=0cm 0cm 0cm 0cm,clip,width=.1\linewidth]{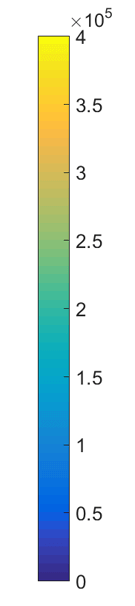}
	\caption{ Stress distribution before and after space-time optimization. Left: After a constrained forward simulation, force residuals are concentrated around constraint points, which leads to an artificial stress distribution. Right: Space-time optimization produces a smoother force residual over the entire spatial domain.}
	\label{fig:sto_stress}
\end{figure}
Space-time optimization is the critical step in the entire pipeline to get training sets from pure kinematic trajectories. For a large scale system or a long trajectory, we need to solve space-time optimization in reduced space. As we can see from the left image of Figure~\ref{fig:sto_stress}, after constrained forward simulation visible nodes are treated as hard constraints, which consequently leads to larger control force or final stress corrections concentrating near visible nodes. This kind of artifact is more obvious when the visible nodes are sparse. It adversely affects the network's ability to learn correct material compensation.
% from the polluted data. 
Through reduced space-time optimization, 
the artificially concentrated force residuals are smoothly distributed in the entire object's domain.

\subsection{Nonlinear Constitutive Material Modelling}

\begin{table*}[t!]
	\begin{center}
		\caption{Statistics measured for different testing cases. From left to right, the test subject, the ground-truth constitutive material model, Young's modulus $E_G$, Poisson ratio $\nu_G$, Rayleigh damping parameters $\alpha_G$ and $\beta_G$ for ground-truth material, nominal material type, the Young's modulus $E_N$, 
			Poisson ratio $\nu_N$, Rayleigh damping parameters $\alpha_N$ and $\beta_N$ used for nominal model. All the Young's modulus values are in MPa, and the size of object is in meters. 
			%for ground-truth materialof nominal material model, Poisson ratio $\nu$ for both nominal and ground-truth model, the size of the subject, maximum position error before training$e_i^t$, maximum position error after training$e_f^t$, maximum position error to test 1$e^{t_1}$, maximum position effor for test 2$e^{t_2}$.
		}
	\label{tab:test_case_statistic}
		\small
		\begin{tabular}{|c|c|c|c|c|c|c|c|c|c|c|c|}
			\hline
			Case	&Size  &Material (GT) & $E_G$ & $\nu_G$ & $\alpha_G$ & $\beta_G$ &Material(N) & $E_N$ & $\nu_N$ & $\alpha_N$ & $\beta_N$   \\
			\hline
			Turtle				& 7x7x3				&NeoHookean   & 2e4  	&0.45  &0.0   &0.0   &Corotation & 3.5e4  & 0.45  &0.0   &0.0  \\
			Dragon 				& 10x4x6			&StVK		  & 1e7 	&0.45  &0.0   &0.0   &Corotation & 7e6    & 0.45  &0.0   &0.0  \\
			Bar (Heterogenous) 	& 0.16x0.16x0.64	&Corotation	  & 1e5/1e7 &0.40  &0.0   &0.0   &Corotation & 3e6    & 0.40  &0.0   &0.0 \\
			Bar1 (Homogeneous) 	& 0.08x0.08x0.32	&Corotation	  & 5e3  	&0.43  &0.0   &0.0   &Corotation & 2.5e3  & 0.43  &0.0   &0.0 \\
			Bar2 (Homogeneous) 	& 0.08x0.08x0.32	&Corotation	  & 5e3  	&0.43  &0.02  &0.1   &Corotation & 2.5e3  & 0.43  &0.0   &0.0 \\
			%Bar3(Homogeneous) 	& 0.08x0.08x0.32	&Corotation	  & 5e3  	&0.43  &0.0   &0.2   &Corotation & 2.5e3  & 0.43  &0.0   &0.0 \\
			\hline
		\end{tabular}
	\end{center}
	\vspace{-15pt}
\end{table*}

\begin{figure}
\includegraphics[trim=2 0cm 0cm 0cm,clip,width=.9\linewidth]{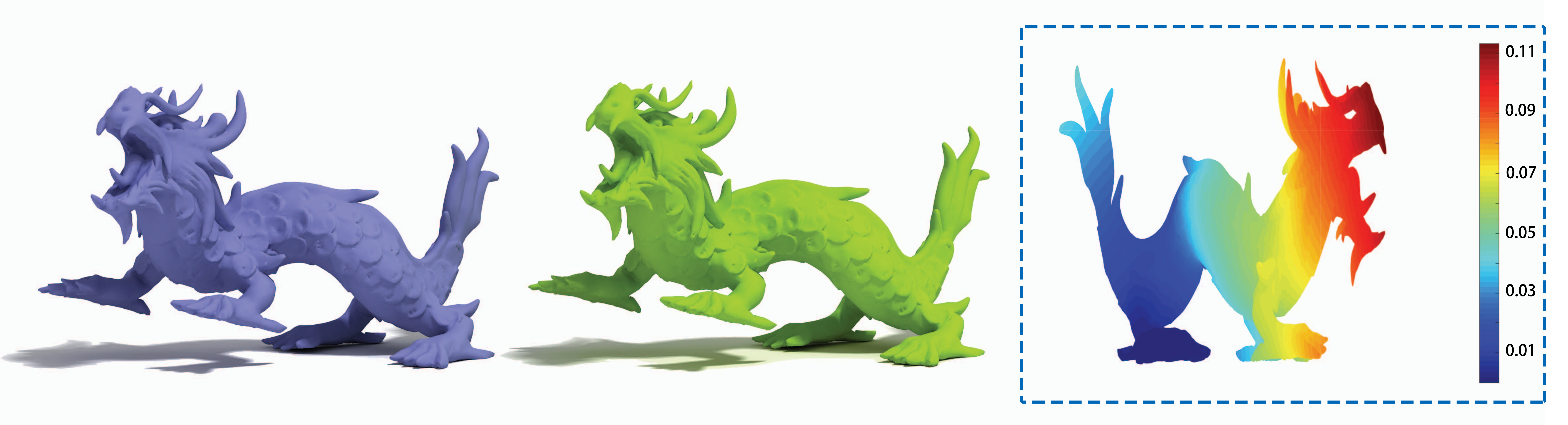}
\caption{Frame with maximum position error in test 2 of the dragon example. The ground truth shape is in purple; the simulation result is in green; 
the rightmost image represents node-wise position error distribution.  } 
\label{fig:Dragon_error_result} 
\end{figure}

\begin{figure}
	\includegraphics[trim=2 0cm 0cm 0cm,clip,width=.9\linewidth]{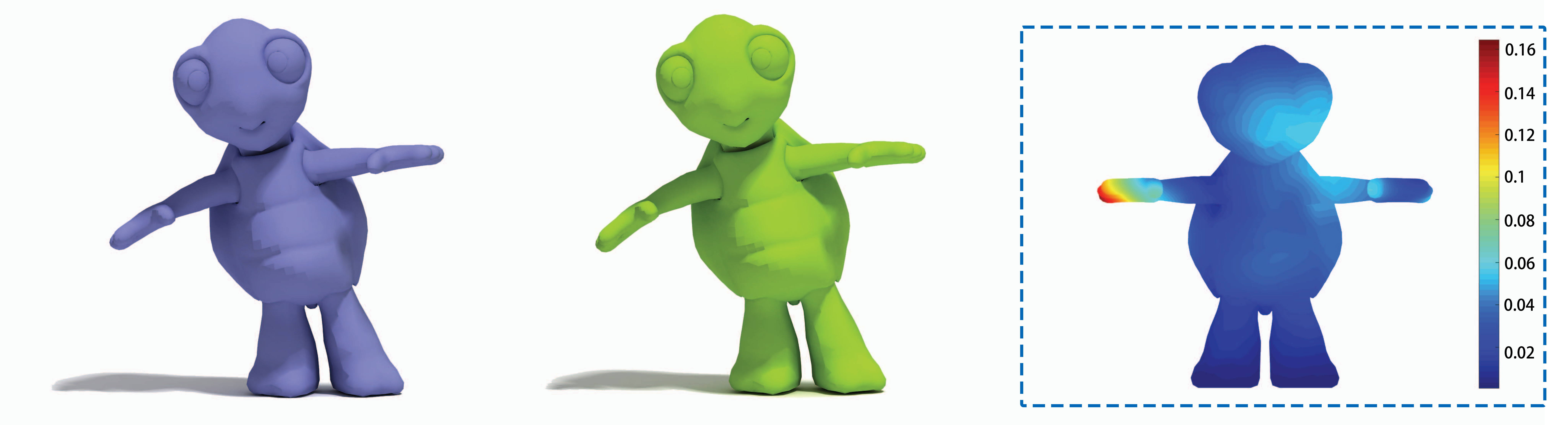}
	\caption{Frame with maximum position error in test 2 of the turtle example. The ground truth shape is in purple; the simulation result is in green; 
the rightmost image represents node-wise position error distribution.  } 
	\label{fig:Turtle_error_result} 
\end{figure}

To validate the generality of our material model estimation algorithm, we generate ground truth trajectories using StVK and Neo-Hookean models in the VEGAFem library without damping, but still keep the nominal one as a co-rotational model. Table~\ref{tab:test_case_statistic} shows the statistics of all our testing cases. The turtle is made of 
Neo-Hokeen material; the dragon is made of StVK material. 
We use two deliberately designed test trajectories to validate our learning result.  The first test has a similar deformation scale as the training trajectory, while second test has
a significantly different range of deformation. As the table and related video show, the learning result can reproduce similar deformation with high accuracy; the result for a different deformation are also satisfying. Vibration differences can only be observed towards the end of a sequence, and as such are due to error accumulation. Figures~\ref{fig:Dragon_error_result} and~\ref{fig:Turtle_error_result} show the frame with maximum position error in the second test for
dragon and turtle examples, respectively.

%The bounding box of the dragon example is observe that the maximum position error on testing cases is typically just a small percentage of the object size.

\subsection{Mateiral Coarsening}
\begin{figure}
	\includegraphics[width=.985\linewidth]{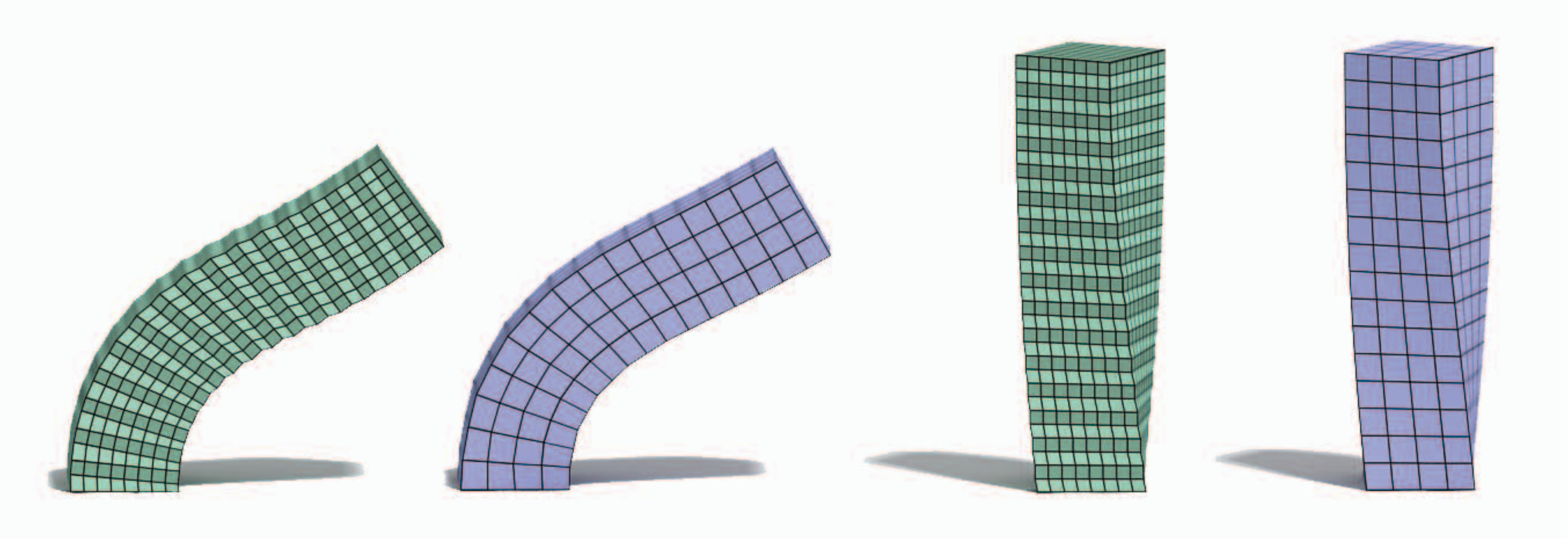}
	\caption{Material coarsening. The green bar represents the fine mesh, with a layered material distribution; 
	the purple bar is the corresponding coarsened mesh, with homogeneous material distribution. Bend (left) and twist (right) deformation trajectories of fine mesh are used as training data, and the purple bars are the reconstruction result after learning. } 
	\label{fig:Coarsening} 
\end{figure} 
The algorithm proposed in this paper can also be used for material coarsening.
%{\color{blue} add reference}.  % PGK: not sure this needs a reference... ?
In Figure~\ref{fig:Coarsening}, a high resolution bar
$(8\times8\times34)$ is composed of two different constitutive materials, with Young modulus values of 1e5 and 1e7, respectively. 
The two materials are composited in a layer by layer manner, represented by the light and dark green colors in Figure~\ref{fig:Coarsening}. The low resolution mesh is the result of coarsening by factor 2 along three axis directions. Two principal deformation modes (bend and twist) are used as training data. The equivalent coarsened material property found by our algorithm can produce very similar motion as the original high resolution heterogeneous model.  

\subsection{Damping Compensation}
To validate the accommodation of our material model estimation algorithm on damping compensation, we use synthetic data generated by forward simulations with known elasticity parameters and damping properties. The Rayleigh damping model is involved when generating the ground truth trajectories. Both the ground truth and nominal material model arise from a co-rotational model, but with different Young's moduli (off by a factor of 2). The first case is a bar model deformed from specific initial configuration with a nonzero stiffness damping coefficient; while the second case has a nonzero mass damping coefficient. The dimension of our bar model is $4\times4\times16$ cm, and as such, we can observe that the maximum position error on testing cases is typically just a small percentage of the object size.
 
\begin{figure}
	\includegraphics[width=\linewidth]{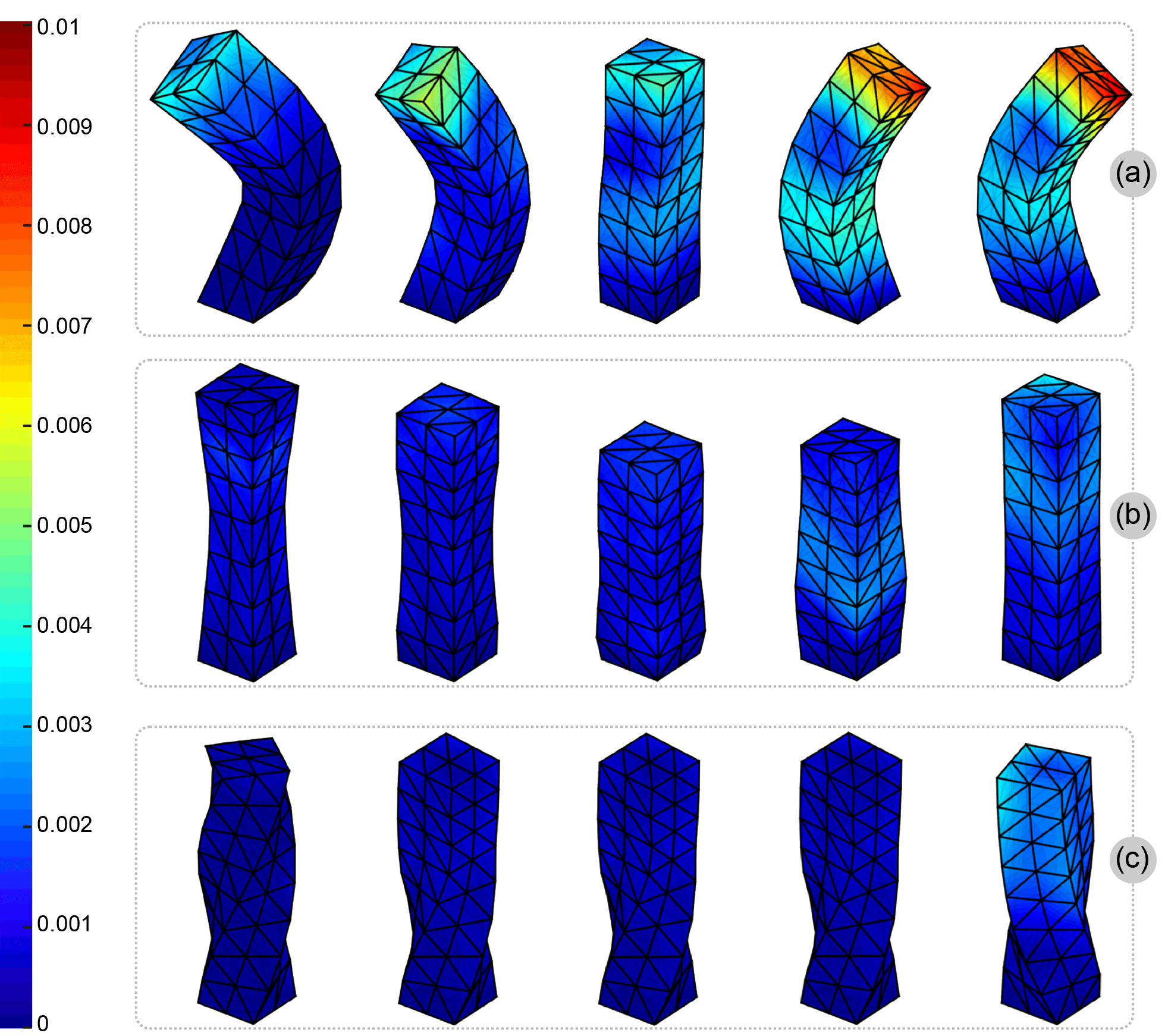}
	\caption{Position error distribution. Position errors are shown from several snapshots of a video (please see supplementary material)
	for different testing scenarios: (a) bend motion with damping; (b) stretch motion without damping; (c) twist motion without damping. 
		%a snapshot from the video showing strain error for different elements. Many of the sequences show no visual error.  The maximum error at any vertex across all frames is in the neighbourhood of 1\% for many examples.  As such, this shows a snapshot from the video showing strain error for different elements. 
	}
	\label{fig:errDist}
\end{figure}

Figure~\ref{fig:errDist} shows snapshots at different times of the error distribution we observe for different examples.  The example involving damping is a challenging case, and we can observe a larger displacement error at different parts of the re-simulated trajectory.  

\begin{table*}[t!]
	\begin{center}
		\caption{Performance statistics measured for different testing cases. Listed from left to right are  the test subject, number of vertices, number of tet elements, number of frames for training data, number of reduced modes, number of learning iterations, maximum position error for training data reconstruction, maximum position error for test 1, maximum position error for test 2, and total computation time for material learning. All the maximum position errors are measured using percentage of object size; the computation times are in hours. }
		\label{tab:accuracy}
		\small
		\begin{tabular}{|c|c|c|c|c|c|c|c|c|c|}
			\hline
			Model (GT)     & \#vert & \#tet & \#frame & \#mode & \#iter  &${err}_L$  &${err}_T^1$  &${err}_T^2$ &$t$ \\
			\hline
			Turtle              &347    &1185   & 600    & 60     & 29       & 3      & 5     & 12  & 6\\
			Dragon              &959    &2590   & 600    & 60     & 33       & 2      & 5     & 8   & 7\\
			Bar (Heterogeneous)    &425    &1536   & 50     & 25     & 41       & 4 / 2  & 4     & -   & 1\\
			Bar1 (Homogeneous)   &81     &192    & 400    & -      & 10       & 1      & 2     & 4   & 1.5\\
			Bar2 (Homogeneous)   &81     &192    & 400    & -      & 49       & 2.5    & 4     & -   & 4\\
			%Bar3(Homogeneous)   &81     &192    & 400    & -      & 49       & 10     & -     & -   & 2\\
			\hline
		\end{tabular}
	\end{center}
	\vspace{-15pt}
\end{table*}

\subsection{Performance}

We measured the computational cost for each critical step on a 10-core 3.0 GHz Intel i7-6950X desktop. 
The performance for space-time optimization, listed in Table~\ref{tab:accuracy}, correlates with the number of tetrahedral elements and the number of frames in the motion trajectory. For our synthesized bar example (192 tetrahedra, 82 vertices, 400 frames), the average computation time for full space-time optimization is 15 minutes per learning iteration. 
When using reduced space-time optimization, the corresponding computation time is approximate 5 minutes. 
%Considering the high dimensionality of the parameter space and the inherent difficulty of the space-time optimization tasks, we deem our optimization algorithm robust and reasonably fast. 
The performance of training is also affected by
%relates with 
the number of hidden nodes: the average training time for 6 hidden nodes is 3 minutes, while 5 minutes were required for 9 hidden nodes.

\section{Conclusion and Future Work}
\label{sec:future}

We have presented a new method called neural material for estimating nonlinear constitutive models from trajectories of surface data.  
The key insight is to have a neural network learn the error of the elastic and damping properties of the material.  
A framework for gradually learning a correction to a nominal material model is described.  The nonlinearity of the force is all in the learning part.
We discuss various neural network designs that can be used for correcting the nominal model, and evaluate training data requirements as well as the necessary number of hidden nodes and layers for successful function approximation.  
Finally, we demonstrate our method with a number of synthetic examples that resemble real world surface capture scenarios.

The desire to work with realistic constitutive models when simulating complex motion has been shared by researchers from many fields, 
not just computer graphics, for a long time. The possibility of employing machine learning technology towards such a goal is tantalizing,
and the present work is a step in that direction. But there is more to be done.
Clearly, the next step for this work is to use scans of real world objects undergoing dynamic motion to estimate neural material models.  
We likewise believe that larger networks that employ six dimensional strain and stress tensors
could be advantageous, though larger networks and larger quantities of example trajectories and training data might be required.  
The damping models we estimate do not currently include hysteresis:  capturing a larger variety of damping behaviours is an important avenue for future research.  
Methods to improve upon the ageing space-time constrains optimization will also be investigated, should our shortcut method prove insufficient.
Finally, for heterogenous materials, there are interesting possibilities for dealing with variation across a model, e.g., by adding an extra input into the neural network to encode a latent material parameter.

\bibliographystyle{ACM-Reference-Format}
\bibliography{DeepM}

\end{document}